\documentclass[
    ,final            % use final for the camera ready runs
  ]
  {aipproc}

\layoutstyle{6x9}

\usepackage{epsfig}
\usepackage{graphicx}
\usepackage{amssymb}

\def\apj{ApJ}%
\def\apjl{ApJ}%
\def\aap{A\&A}%
\def\mnras{MNRAS}%
\def\nat{Nature}

\def\msun{{\rm M_\odot}}

\def\Te{T_{\rm e}}

\def\rg{R_{S}}
\def\betaeq{\beta_{\rm eq}}
\def\Dop{\delta}
\def\betaeq{\beta_{\rm eq}}
\def\taut{\tau_{\rm T}}
\def\degr{^{\rm o}}

\def\sax1808{SAX~J1808.4--3658}
\def\second{XTE~J1751--305}

\def\fifth{XTE~J1814--338}
\def\igr{IGR~J00291+5934}

\newcommand{\be}{\begin{equation}}
\newcommand{\ee}{\end{equation}}
\newcommand{\beq}{\begin{eqnarray}}
\newcommand{\eeq}{\end{eqnarray}}
\newcommand{\beqno}{\begin{eqnarray*}}
\newcommand{\eeqno}{\end{eqnarray*}}
\newcommand{\bitmz}{\begin{itemize}}
\newcommand{\eitmz}{\end{itemize}}

\begin{document}

\title{Modeling pulse profiles \\ of accreting millisecond pulsars}


\author{Juri Poutanen}{address={Astronomy Division, Department of Physical Sciences, 
P.O. Box 3000,   FIN-90014 University of Oulu, Finland}}

\begin{abstract}
I review the basic observational properties of accreting millisecond pulsars that are important for understanding the physics involved in formation of their pulse profiles. I then discuss main effects responsible for shaping these profiles. Some analytical results that help to understand the results of simulations are presented. Constraints on the pulsar geometry and the neutron star equation of state obtained from the analysis of the pulse profiles are discussed.  
\end{abstract}

\maketitle

%%%%%%%%%%%%%%%%%%%%%%%%%%%%%%%%%%%%%%%%%%%%
%% MAINMATTER
%%%%%%%%%%%%%%%%%%%%%%%%%%%%%%%%%%%%%%%%%%%%

\section{Introduction}

First evidences for  rapid rotation of neutron stars (NS) in low-mass X-ray binaries appeared in 1996 with the discoveries of sub-kHz coherent oscillations observed during X-ray bursts by the {\it Rossi X-ray Timing Explorer (RXTE)} \cite[see][ for review]{SB06}. Two years later, the first accreting X-ray millisecond pulsar (AMP) was discovered \cite{WvdK98}, which was a nice confirmation of the theory of production of millisecond radio pulsar in the course of accretion. By 2008, there are at least 8 accreting NS discovered that show coherent oscillations for the extended periods of time during the outbursts \cite{W06,P06}. These sources are exceptional laboratories to study the physics of accretion onto magnetized stars. They also have a great potential to test the NS structure. These goals can be achieved by studying the pulse profiles.  This, however, requires understanding of the processes responsible for the production of the X-rays at the NS surface  as well as detailed modeling of the propagation of the radiation to the observer. 

I review here the main observational data that have direct relation to this topic. These include averaged and phase-resolved broad-band spectra, pulse profiles and their energy dependence, and time lags. Then I describe recent efforts to develop theoretical models devoted to modeling these data and the main results obtained from these studies. 

\section{Observations}

\subsection{Broad-band spectra}

\begin{figure}[h]
%\centerline{\leavevmode \epsfxsize=7.0cm \epsfbox{fig_sax1808.ps}  \epsfxsize=7.0cm \epsfbox{fig_xte1751.ps}}
  \includegraphics[height=.23\textheight]{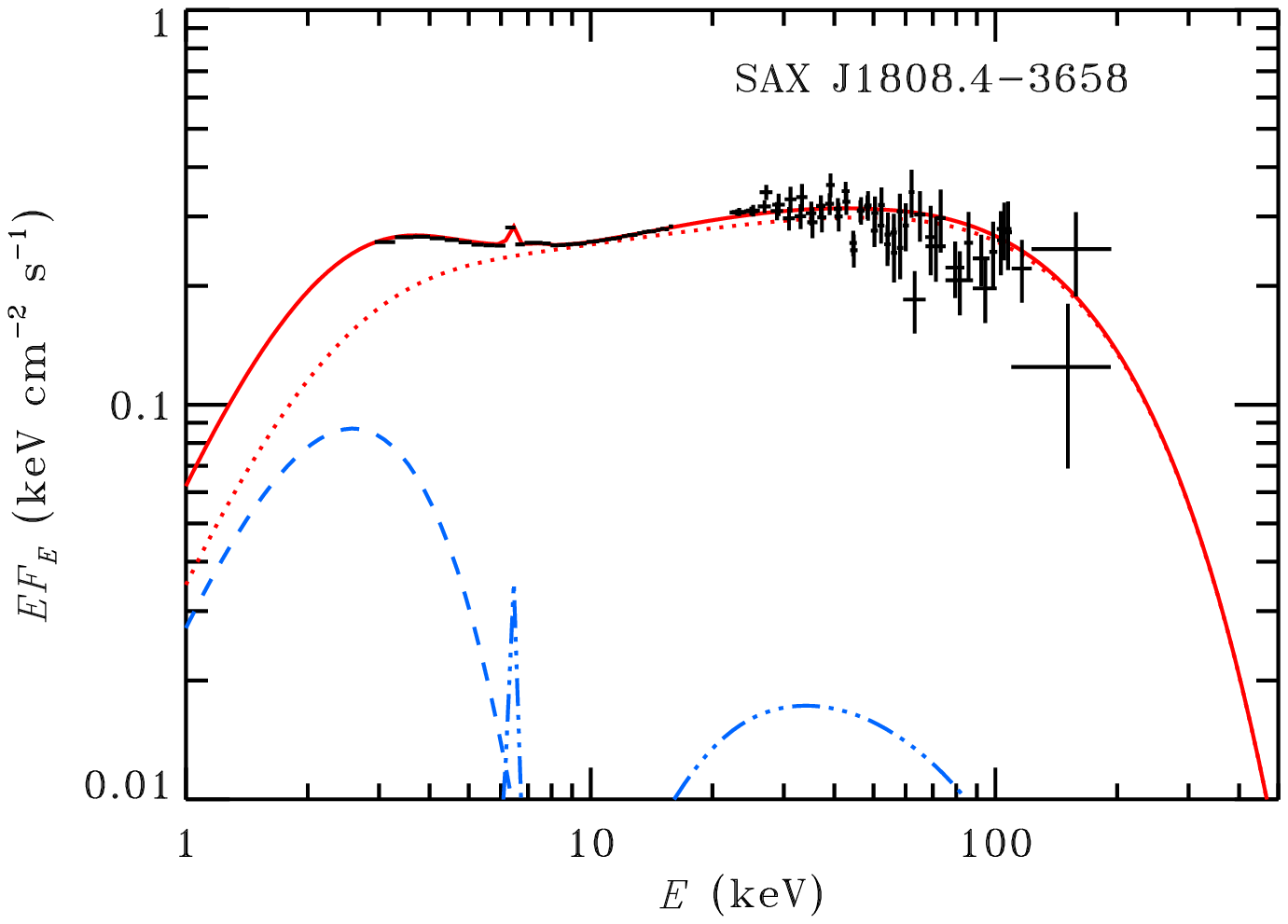}
    \includegraphics[height=.23\textheight]{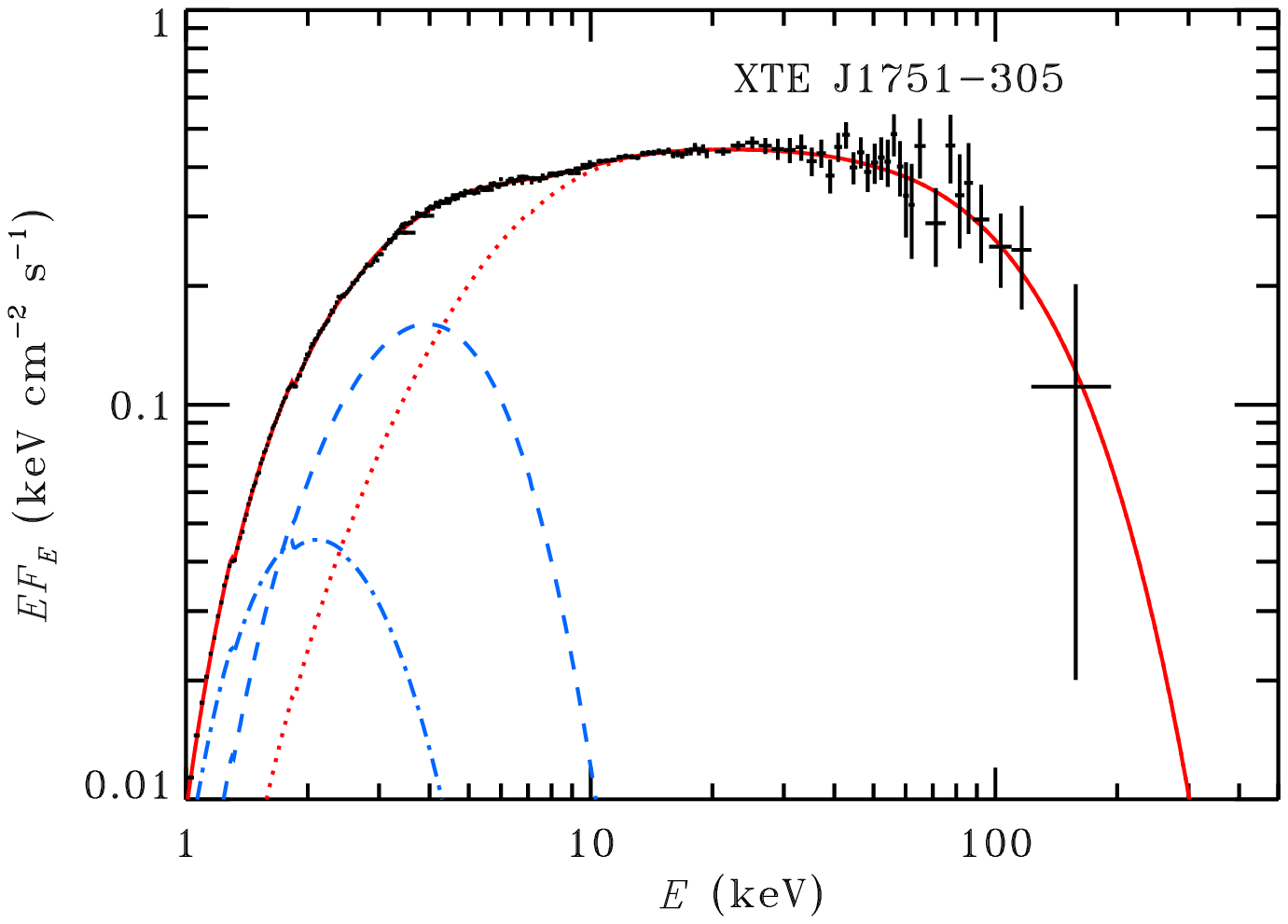}
  \caption{Broad-band spectra of SAX J1808.4-3658 (left) and XTE J1751-305 (right). 
  Dotted and dashed curves show the Comptonized and blackbody components, respectively. The reflection 
  is shown by  triple-dot-dashed curve and an additional soft disk component is shown by dot-dashed curve.   }
  \label{fig:spectra}
\end{figure}

In order to understand how the X-ray pulses are produced in AMP, we need first to identify the basic radiation processes responsible for the X-rays we observe. The broad-band coverage of the {\it RXTE} together with {\it XMM-Newton} and {\it INTEGRAL} gives a possibility to study the spectra of AMP in great details. The AMP turn out to have rather similar, power-law like spectra with photon index $\Gamma\sim1.8-2.1$ with a cutoff around 100 keV \cite{GR98,GDB02,PG03,GP05, FBP05,FKP05,FPB07}  (see Fig.~\ref{fig:spectra}).\footnote{Spectral fits of the {\it Chandra} or {\it XMM-Newton} data alone in a narrow energy band give much harder  slopes with $\Gamma\sim$1.4--1.5 \cite[see e.g.][]{JGC03,MWM03}.} Such spectral shape strongly argues in favor of thermal Comptonization (with electron temperature $k\Te\sim 50$ keV and Thomson optical depth of $\taut\sim1$) as the main emission process, while the absence of strong spectral variability can be used as an argument that the emission region geometry does not depend much on the accretion rate and the magnetic field strength. The seed photons for Comptonization are probably produced by reprocessing of the hard radiation at the NS surface (similarly to the two-phase model developed for Seyfert galaxies \cite{HM93,PS96,MBP01}), and not cyclo-synchrotron radiation. Below $\sim$5 keV, a soft excess is often present. It can be associated with the blackbody like emission  (with $kT \sim 0.4-0.7$ keV)  from the heated NS surface, but not the accretion disk, because it is pulsating.

\begin{figure}
%\centerline{\epsfig{file=fig_sax1808_twocomp.ps,width=12truecm}}
  \includegraphics[height=.25\textheight]{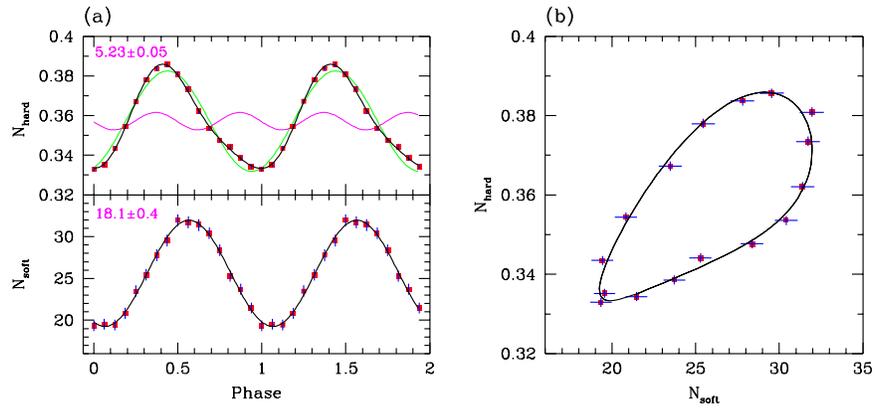}
\caption{Pulsation  of the two model components, blackbody and Comptonization, in \sax1808; from \cite{GDB02}. 
Similar behavior is seen in \second, see \cite{GP05}.}
\label{fig:fars_bbtf}
\end{figure}

It is also of interest to study spectral variability as a function of pulsar phase. The simplest model has only two parameters corresponding to the normalizations of the two components (blackbody + Comptonization). One sees a hysteresis-like loop  (see Fig. \ref{fig:fars_bbtf}), with the  blackbody lagging the Comptonized emission \cite{GDB02,GP05}. This behavior  results in the soft time-lags (discussed below). The oscillation amplitude and the pulse shape of the two components are also significantly different.

\subsection{Pulse profiles}

Most of the pulsars show extremely simple pulse profiles, which can well be described by just one sine wave. However, at high energies above 10 keV contribution of the harmonic becomes stronger (with the exception of \igr, \cite{FKP05}), the profile becomes more skewed with faster rise and slower decay.  During the decay of the 2002 outburst of \sax1808, the profile show double peaks, which is probably the evidence for appearance of the secondary spot. Strong profile changes also result in a very noisy behavior of the pulsar phase \cite{BDM06,RDB08}.

\begin{figure}
 %\centerline{\epsfig{file= fig_sax1808_at_2002.ps,width= 10truecm}}
  \includegraphics[height=.4\textheight]{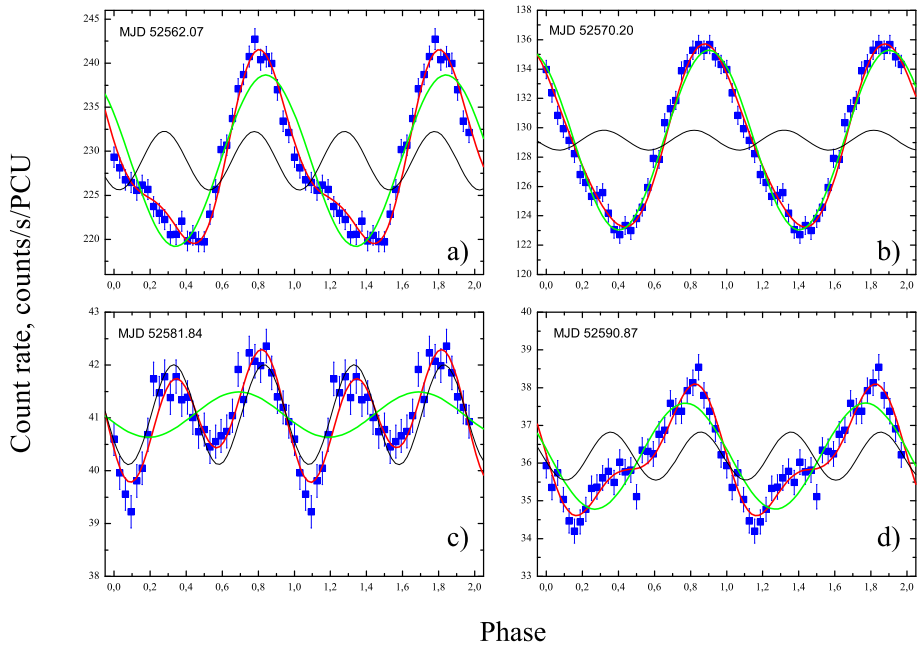}
  \caption{Pulse profiles observed from \sax1808\ during its 2002 outburst and 
  the best-fit sinewaves. (From Ibragimov et al., in preparation.)}
  \label{fig:sax_at_2002}
\end{figure}

\subsection{Time lags}

Pulse profiles at higher energies reach their peaks at an earlier phase relative to the soft photons resulting in the soft time lags \cite{CMT98,GDB02, GP05, GMK07,Chou08}, with the absolute value increasing with energy up to about 7--10 keV and saturating afterwards. Only in \igr, the trend seem  to reverse above 10 keV \cite{GM05,FKP05}. Here it is important to note that the blackbody's contribution   decreases exponentially with energy and around 7 keV it is negligible.  Thus, changes in the pulse profile of the two components are the obvious cause of the lags. One cannot separate lags from the pulse profiles and any successful time-lag model should actually explain not the lags, but the difference in the pulse profiles of the  spectral components. A difference in  the  emission pattern \cite{PG03} can be responsible for that. On the other hand, models involving Compton down- (or up-) scattering \cite{CMT98,FT07} pay attention only to the lags ignoring the pulse profiles.\footnote{A specific time-lag model of \cite{FT07} suffers also from other problems. There the soft lags are explained by multiply scattering of the hotspot's hard radiation in the accretion disk. However, the radiation above 10 keV is actually mostly reflected, while the few keV photons are immediately absorbed as the disk is rather cool. Thus, the role of multiply scattering is negligible and the lags cannot possibly be produced this way.}

\section{Theory of pulse profiles}

The observed pulse profiles depend on a number of parameters: 
\begin{itemize}
\item Geometrical parameters, inclination $i$ and co-latitude $\theta$ of the magnetic poles. 
\item Geometry of the accretion shock and the hot spot at the NS surface, which in turn depend on the 
accretion rate $\dot{M}$, magnetic field strength  and $\theta$. 
\item Emissivity pattern of the radiation, which may depend on $\dot{M}$. 
\item Stellar spin $\nu$, which affects the Doppler boosting and aberration effects. The importance of time delays, related to the finite propagation time around a NS,   also depend on $\nu$.
\item Stellar compactness, i.e. NS radius-to-mass ratio $R/M$, which has a dramatic effect on the strength of gravitational bending.  
\item Stellar oblateness, which might become important for stars of large radii rotating at $\nu\gtrsim 600$ Hz.
\item Accretion disk inner radius, which blocks radiation from the antipodal spot and also reflects hard radiation.  
\item Accretion stream's optical depth, which depends on $\dot{M}$. The stream can eclipse the hotspot at some pulsar phases. 
\end{itemize}
The quality of the data does not justify yet to account for all possible effects. Many models make serious simplifications, for example, considering  circular spots, spherical stars, blackbody emission, etc.  

\subsection{Main effects}

\begin{figure}
%\centerline{\epsfig{file= fig_relat_effects.ps,width= 10truecm}}
  \includegraphics[height=.32\textheight]{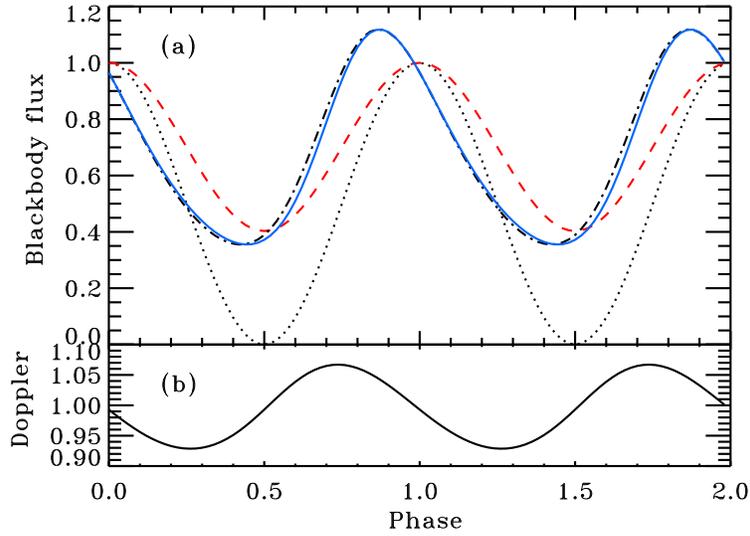}
   \caption{Pulse profile for a small blackbody spot   at the neutron star surface.
  (a) Dotted curve shows the  profile for a   slowly rotating star in Newtonian approximation.
  Gravitational light bending reduces variability amplitude (dashed curve).   Doppler effect due to rapid rotation  skews the profile
  (dot-dashed curves). Accounting for the light travel delays   further modifies the profile slightly (solid curve). (b) Doppler factor
  as a function of phase.   We assumed neutron star mass $M=1.4\msun$ and radius  $R=2.5\rg=10.3$ km ($\rg=2GM/c^2$), rotational frequency $\nu=600$ Hz,  inclination $i=45^{\rm o}$, and   the colatitude of the spot center $\theta=45^{\rm o}$. From \cite{P06}. }
\label{fig:relat}
\end{figure}

Let us now briefly describe the main effects shaping the pulse profiles of AMP. Assume a small blackbody spot. Without general or special relativistic effects, a spot would produce sinusoidal variations (with possible eclipses) due to a change of the projected visible area. The gravitational light bending tends to decrease the pulse amplitude \cite{PFC83,RM88,Page95, LL95,ZSP95,B02} (compare dotted and dashed curves in Fig.~\ref{fig:relat}),  while the pulse remains almost sinusoidal \cite{B02}.  The importance of Doppler boosting and  light travel time delays was pointed out in \cite{BRR00}, who also showed that  the frame dragging affects the pulse profiles only at a $\sim1\%$ level. 
Therefore, one can use Schwarzschild metric. To account for the Doppler effect one makes Lorentz transformation from the frame rotating with the star to a non-rotating frame and then follows the light trajectory in  the Schwarzschild space-time to infinity (``Schwarzschild + Doppler" or SD approach).  The full formalism for computing the pulse profiles accounting for Doppler boosting, relativistic aberration and gravitational bending appeared for the first time in \cite{PG03} (see \cite{PB06} for detailed derivations). Accuracy of the SD approach is discussed in \cite{CLM05}. Comparing to the slowly rotating stars, the projected area is changed by the Doppler factor $\Dop$ due to aberration, and the bolometric intensity is multiplied by $\Dop^4$ \cite[see][]{PG03,PB06}. Since $\Dop$ reaches the maximum  a quarter of the period earlier than the projected area, the pulse becomes skewed to the left (see Fig.~\ref{fig:relat}). Light travel time delays  slightly modify the profile further. Oscillation waveforms and amplitudes were computed in SD approach in \cite{WML01,MOC02}, who however present no formulae. The graphs shown in \cite{WML01,MOC02} can be understood easily using analytical formulae for oscillation amplitudes derived in \cite{P04,PB06} (see Fig. \ref{fig:ampl}). 
   
The light curves from realistic spots produced by accretion onto inclined magnetic dipole are computed in \cite{KR05} using formalism of \cite{PG03}. Also stellar oblateness might affect the pulse profiles for rapidly rotating stars \cite{CMLC07,MLCB07}. Effects of the anisotropy of Comptonized radiation from the shock on the light curves and polarization   were studied in \cite{VP04}. 

\subsection{Angular pattern}

The angular pattern of the radiation escaping the NS surface plays an important role in shaping the pulse profile. The shape of the broad-band spectra of AMP implies that Comptonization happens in the optically thin atmosphere.  So far only  static atmospheres of constant temperatures have been considered \cite{PG03,VP04}. As blackbody photons from the NS surface are scattered in the hot shock, they gain energy.  The angular distribution of photons strongly depends on the number of scattering (and therefore their energy), saturating after about 5 scatterings. Radiative transfer calculations show that the angular dependence of the specific intensity can be approximated by a linear relation $I(\mu)\propto 1+a\mu$ (where $\mu=\cos\alpha$ and $\alpha$ is the angle between photon propagation direction and the local normal). For optically thin shocks, the scattered radiation correspond to negative $a\lesssim-0.5$.  This gives pulse profiles dramatically different from those obtained assuming the blackbody (i.e. $a=0$) or the Hopf profile (i.e. $a=2$) corresponding to  the optically thick electron scattering atmosphere \cite{Cha60,Sob63}. Changes in the emission pattern happen smoothly with energy until blackbody flux becomes negligible (above 7--10 keV, see Fig. \ref{fig:spectra}) and only scattered radiation is  left. Correspondently, the pulse profile shape and the position of their maxima moves to earlier phase \cite{PG03}. This results in the soft time lags as observed. Thus Comptonization model explains both the pulse profiles and the time lags. I note that the lags due to finite travel time in the Comptonizing medium are completely irrelevant here.

\subsection{Analytical approximations}

\begin{figure}
 %\centerline{\epsfig{file= fig_ampl.ps,height= 6.5truecm}}
 \includegraphics[height=.3\textheight]{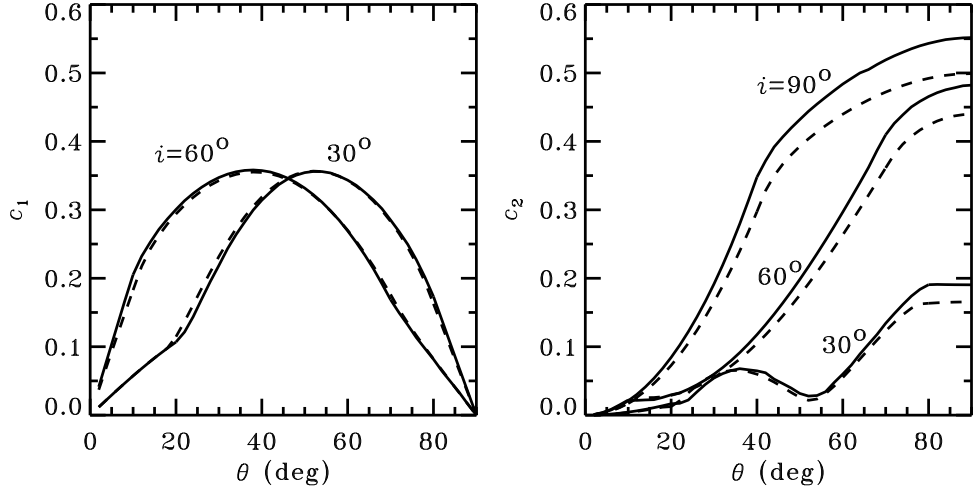}
  \caption{Amplitudes of the fundamental and the harmonic for the pulsars with two antipodal spots as a function of magnetic colatitude $\theta$. 
  Note that for $i=90\degr$, only a harmonic is present.   Solid curves are exact calculations and dashed curves are analytical approximation for $M=1.4\msun$, $R=2.5\rg$, $\nu=600$ Hz,   and $a=-0.7$. Adapted from \cite{PB06}. }
 \label{fig:ampl} 
\end{figure}

As the pulse profiles depend in a complicated way on the number of parameters, it is easier to understand their behavior using analytical approximations. In an important paper,  \citet{B02} showed that in Schwarzschild metric there is a simple relation between the angle the photon momentum makes with the normal $\alpha$ and that angle without the light bending $\psi$ \citep[see also][]{ZSP95}: 
\be\label{eq:cosbend}
\cos\alpha\approx u + (1-u) \cos\psi , 
\ee
where $u=\rg/R$,  $\cos\psi=\cos i\ \cos\theta+\sin i \sin\theta\cos\phi$,  and $\phi$ is the pulsar phase. With the same accuracy the time delays can be computed using a relation $\Delta t=(1-\cos\psi)R/c$ and the Doppler factor  can also be approximated as \cite{PB06}
\be
\delta = 1- \betaeq\sqrt{1-u}\ \sin i\ \sin \theta \cos\phi, 
\ee
where $\betaeq=v_{\rm eq}/c=2\pi R \nu/c\sqrt{1-u}$ is the equatorial velocity. 
This allows to compute the pulse profiles, Fourier amplitudes and phases analytically with high accuracy (see Fig. \ref{fig:ampl}).

Analytical formulae allow to understand easily that for one spot the pulse amplitude increases linearly with $\sin i\ \sin \theta$. The deviation of the pulse profile from the sine-wave (i.e. the ratio of the harmonic to fundamental) also depends linearly on $\sin i\ \sin \theta$, the rotational velocity $\betaeq\ll 1$ as well as  on  the anisotropy  parameter $a$ (which does not need to be small).  Thus strong harmonics (skewness of the profile) are mostly a result of the emission anisotropy. We expect strong harmonics only if the pulse itself has a large amplitude (as e.g. in \fifth, \cite{SM03}). 

It is known that, for the blackbody emitting spots, oscillations almost disappear when both spots are visible (for the so called class IV pulsars, \cite{B02}), because each spot produces a sine wave which are shifted in phase by $180\degr$. An anisotropic source with $a\neq 0$, however, produces strong oscillations. The amplitudes of the fundamental  and harmonic behave roughly as  $\propto a\ \sin 2i\ \sin 2\theta$ and $\propto a\ \sin^2 i\ \sin^2\theta$, respectively \cite{PB06}, with a weak dependence on stellar compactness and rotational frequency.  It is clear thus that even a slight anisotropy is capable of producing strong pulsations.  Stellar rotation alone, even for $a=0$, also produces oscillations with amplitudes  $\propto \betaeq\ \sin 2i\ \sin 2\theta$ and $\propto \betaeq\ \sin^2i\ \sin^2\theta$. This argues in favor of weak magnetic field (and not their large mass as proposed recently in \cite{ozel08}) in the majority of NS in low-mass X-ray binaries as the main reason for rarity  of AMP. Alternatively, a nearly perfect alignment of the magnetic and rotational axes, $\theta\sim 0$, solves the problem \cite{lamb08}.  
 
\begin{figure} 
%\centerline{\epsfig{file= fig_itheta_3rg.ps,width= 7.0truecm}}
 \includegraphics[height=.3\textheight]{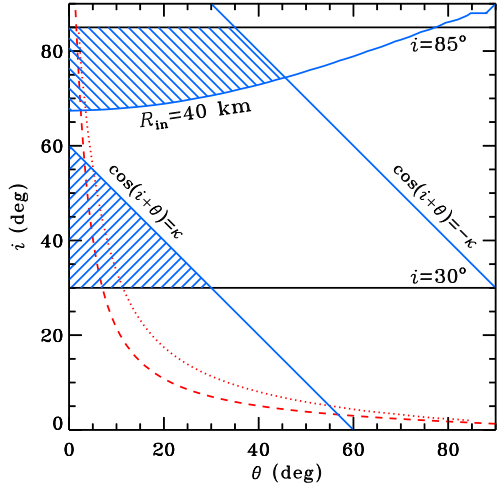}
\caption{Constraints on the inclination and magnetic inclination for \second. The lines $i=30\degr$ and $i=85\degr$ are the lower and upper limits on the inclination of the system \cite{MS02}. The absence of eclipses of the primary spot limits the allowed region to $\cos(i+\theta)>-\kappa$ (where $\kappa\equiv R/(R-\rg)$;  to the left from the corresponding line, see \cite{B02} for description of the regions). The antipodal spot can be invisible if it is blocked by the star, this gives  the allowed regions of parameters $i$ and $\theta$,  $\cos(i+\theta)>\kappa$ (lower hatched region). It can also be blocked by the accretion disk (inner disk radius of $R_{\rm in}=40$ km is assumed), this gives another allowed region (upper hatched area). 
Contours of constant variability amplitude $A=0.045$ are shown for the blackbody spot at a slowly rotating star (dashed curve) and for the  Comptonized radiation (dotted curve) multiply scattered in the shock of Thomson optical depth $\taut=1.5$ and a star rotating at $\nu=435$ Hz (as in \second). NS of mass $M=1.4 \msun$ and radius $R=3\rg$ is assumed. From \cite{GP05}. }
\label{fig:itheta}
\end{figure}

Analytical approach gives  the peak-to-peak amplitude (for $a\geq -1/2$, small spot and slow rotation) of nearly sinusoidal pulses \cite{PB06}
\be \label{eq:}
A\equiv \frac{F_{\max}-F_{\min}}{F_{\max}+F_{\min}} = \frac{U(1+2aQ)}{Q+a(Q^2+U^2)} ,
\ee
where $Q=u+(1-u)\cos i \cos\theta$, $U=(1-u)\sin i\sin\theta$. This simple relation allows us to understand the general behavior of the oscillation amplitude even for the rapid rotation, which can be used to put constraints on $i$ and $\theta$  (see Fig. \ref{fig:itheta}).

\subsection{Accretion disk: eclipses and reflection}

As the inner edge of the accretion disk in AMP reaches  a few stellar radii, the secondary accretion spot is invisible for almost any inclination. This is supported by the nearly sinusoidal pulse profiles observed in most AMP. At low accretion rate, the Alfv\'en radius grows and the disk retreats towards the corotation radius allowing direct view of the antipodal spot.  The profile becomes much more complicated (as seen in \sax1808, Fig. \ref{fig:sax_at_2002}), which results in the phase jumps of the fitted sinusoids. This possibly explains a complicated behavior of the pulsar phase \cite{BDM06,RDB08}. 

Using simple analytical formulae for light bending derived in \cite{B02}, one can get constraints on the inclination and magnetic angle $\theta$ from the absence of the secondary spot \cite{GP05}.  It can be blocked either by the accretion disk implying rather high inclination (see upper hatched region in Fig. \ref{fig:itheta}) or by the star itself (lower hatched region). However, if double peaked profiles are sometimes seen (as is the case for \sax1808\ at low luminosity, see Fig. \ref{fig:sax_at_2002}), only the first solution is possible. 

The accretion disk not only absorbs some radiation, but also reflects hard X-rays. The reflection amplitude depends on the inner disk radius and stellar compactness \cite{GP05} (as well as on the emission pattern and spot size and position). Variation of the reflection with the accretion rate  can be used to track the position of the inner disk radius.    

 \subsection{Constraints on the NS equation of state}

Detailed fitting of the AMP pulse profiles allows us to constrain the NS equation of state. Analysis of the data on \sax1808\   resulted in constraints on the NS radius $8<R<11$ km for $M=1.4$--$1.6\msun$ \cite{PG03}. Recently, these results were confirmed with the model including also  time delays and stellar oblateness  \cite{LMC08}.  However, a study of \fifth\ \cite{LMCC08} gave a large NS radius requiring a stiff equation of state. This contradiction possibly originates either in the assumption of the spot geometry and/or simplicity of the emission model. Further studies on how these assumptions affect the results are  required. 

\section{Conclusions}

I have reviewed the basic observational properties of AMP such as spectral energy distribution, pulse profiles, time lags that are important for understanding of the physics of accretion in these sources.  I also discussed the major effects that are responsible for shaping the pulses in AMP: light bending, Doppler boosting, anisotropy of emission, eclipses by the accretion disk, etc. The complicated dependences of the pulses on various parameters trigger the analytical approach which allows to understand better the results of simulations. I discussed how various observables such as amplitudes of fundamental and harmonic depend on the main parameters as well as how the invisibility of the secondary spot  constrains geometrical parameters and the inner disk radius. This in turn can be used to constrain the pulsar magnetic field. Various attempts to measure NS compactness produce contradicting results, which probably reflects our ignorance of the angular distribution of the radiation as well as the spot geometry.  Theoretical studies of these aspects of accretion onto AMP are certainly welcome.

 %%%%%%%%%%%%%%%%%%%%%%%%%%%%%%%%%%%%%%%%%%%%%%%%
%% BACKMATTER
%%%%%%%%%%%%%%%%%%%%%%%%%%%%%%%%%%%%%%%%%%%%%%%%

\begin{theacknowledgments}
I acknowledge the support from the Academy of Finland grant 110792. 
I thank my collaborators on this topic Andrei Beloborodov, Maurizio Falanga, Marek Gierli\'nski, and Askar Ibragimov. 
\end{theacknowledgments}



\bibliographystyle{aipprocl} % if natbib is missing

%\bibliography{ns}

\begin{thebibliography}{47}
\expandafter\ifx\csname natexlab\endcsname\relax\def\natexlab#1{#1}\fi
\providecommand{\enquote}[1]{``#1''}
\expandafter\ifx\csname url\endcsname\relax
  \def\url#1{\texttt{#1}}\fi
\expandafter\ifx\csname urlprefix\endcsname\relax\def\urlprefix{URL }\fi
\providecommand{\eprint}[2][]{\url{#2}}

\bibitem[{Strohmayer} and {Bildsten}(2006)]{SB06}
T.~{Strohmayer}, and L.~{Bildsten}, 
%\enquote{{New views of thermonuclear bursts},} 
% Cambridge Astrophysics Series, No. 39
 in \emph{Compact stellar X-ray sources}, edited by W.~{Lewin}, and M.~{van der Klis}, Cambridge
  University Press, Cambridge, 2006, pp. 113--156.

\bibitem[{Wijnands} and {van der Klis}(1998)]{WvdK98}
R.~{Wijnands}, and M.~{van der Klis}, \emph{\nat} \textbf{394}, 344--346
  (1998).

\bibitem[{Wijnands}(2006)]{W06}
R.~{Wijnands}, 
%\enquote{{Accretion-driven millisecond X-ray pulsars},} 
in \emph{Trends in Pulsar Research}, edited by J.~A. {Lowry}, Nova Science
  Publishers, New York, 2006, pp. 53--78, \eprint{arXiv:astro-ph/0501264}.

\bibitem[{Poutanen}(2006)]{P06}
J.~{Poutanen}, \emph{Advances in Space Research} \textbf{38}, 2697--2703
  (2006). %, \eprint{arXiv:astro-ph/0510038}.

\bibitem[{Gilfanov} et~al.(1998)]{GR98}
M.~{Gilfanov}, M.~{Revnivtsev}, R.~{Sunyaev}, and E.~{Churazov}, \emph{\aap}
  \textbf{338}, L83--L86 (1998).

\bibitem[{Gierli{\'n}ski} et~al.(2002)]{GDB02}
M.~{Gierli{\'n}ski}, C.~{Done}, and D.~{Barret}, \emph{\mnras} \textbf{331},
  141--153 (2002).

\bibitem[{Poutanen} and {Gierli{\'n}ski}(2003)]{PG03}
J.~{Poutanen}, and M.~{Gierli{\'n}ski}, \emph{\mnras} \textbf{343}, 1301--1311
  (2003).

\bibitem[{Gierli{\'n}ski} and {Poutanen}(2005)]{GP05}
M.~{Gierli{\'n}ski}, and J.~{Poutanen}, \emph{\mnras} \textbf{359}, 1261--1276
  (2005).

\bibitem[{Falanga} et~al.(2005{\natexlab{a}})]{FBP05}
M.~{Falanga}, J.~M. {Bonnet-Bidaud}, J.~{Poutanen}, et al., 
%R.~{Farinelli}, A.~{Martocchia}, P.~{Goldoni}, J.~L. {Qu}, L.~{Kuiper}, and A.~{Goldwurm},
  \emph{\aap} \textbf{436}, 647--652 (2005{\natexlab{a}}). %,   \eprint{arXiv:astro-ph/0503292}.

\bibitem[{Falanga} et~al.(2005{\natexlab{b}})]{FKP05}
M.~{Falanga}, L.~{Kuiper}, J.~{Poutanen}, et al., 
%E.~W. {Bonning}, W.~{Hermsen}, T.~{di Salvo}, P.~{Goldoni}, A.~{Goldwurm}, S.~E. {Shaw}, and L.~{Stella},
  \emph{\aap} \textbf{444}, 15--24 (2005{\natexlab{b}}). %,   \eprint{arXiv:astro-ph/0508613}.

\bibitem[{Falanga} et~al.(2007)]{FPB07}
M.~{Falanga}, J.~{Poutanen}, E.~W. {Bonning}, et al. 
%L.~{Kuiper}, J.~M. {Bonnet-Bidaud}, A.~{Goldwurm}, W.~{Hermsen}, and L.~{Stella}, 
\emph{\aap} \textbf{464}, 1069--1074 (2007).% , \eprint{arXiv:astro-ph/0609776}.

\bibitem[{Juett} et~al.(2003)]{JGC03}
A.~M. {Juett}, D.~K. {Galloway}, and D.~{Chakrabarty}, \emph{\apj}
  \textbf{587}, 754--760 (2003). %, \eprint{arXiv:astro-ph/0208543}.

\bibitem[{Miller} et~al.(2003)]{MWM03}
J.~M. {Miller}, R.~{Wijnands}, M.~{M{\'e}ndez}, et al., 
%E.~{Kendziorra}, A.~{Tiengo}, M.~{van der Klis}, D.~{Chakrabarty}, B.~M. {Gaensler}, and W.~H.~G. {Lewin},
  \emph{\apjl} \textbf{583}, L99--L102 (2003). %, \eprint{arXiv:astro-ph/0208166}.

\bibitem[{Haardt} and {Maraschi}(1993)]{HM93}
F.~{Haardt}, and L.~{Maraschi}, \emph{\apj} \textbf{413}, 507--517 (1993).

\bibitem[{Poutanen} and {Svensson}(1996)]{PS96}
J.~{Poutanen}, and R.~{Svensson}, \emph{\apj} \textbf{470}, 249--268 (1996).

\bibitem[{Malzac} et~al.(2001)]{MBP01}
J.~{Malzac}, A.~M. {Beloborodov}, and J.~{Poutanen}, \emph{\mnras}
  \textbf{326}, 417--427 (2001).

\bibitem[{Burderi} et~al.(2006)]{BDM06}
L.~{Burderi}, T.~{Di Salvo}, M.~T. {Menna}, A.~{Riggio}, and A.~{Papitto},
  \emph{\apjl} \textbf{653}, L133--L136 (2006). %,   \eprint{arXiv:astro-ph/0612093}.

\bibitem[{Riggio} et~al.(2008)]{RDB08}
A.~{Riggio}, T.~{Di Salvo}, L.~{Burderi}, et al., 
% M.~T. {Menna}, A.~{Papitto}, R.~{Iaria}, and G.~{Lavagetto}, 
\emph{\apj} \textbf{678}, 1273--1278 (2008). %,  \eprint{arXiv:0710.3450}.

\bibitem[{Cui} et~al.(1998)]{CMT98}
W.~{Cui}, E.~H. {Morgan}, and L.~G. {Titarchuk}, \emph{\apjl} \textbf{504},
  L27--L30 (1998).

\bibitem[{Galloway} et~al.(2007)]{GMK07}
D.~K. {Galloway}, E.~H. {Morgan}, M.~I. {Krauss}, P.~{Kaaret}, and
  D.~{Chakrabarty}, \emph{\apjl} \textbf{654}, L73--L76 (2007). %,  \eprint{arXiv:astro-ph/0609693}.

\bibitem[{Chou} et~al.(2008)]{Chou08}
Y.~{Chou}, Y.~{Chung}, C.-P. {Hu}, and T.-C. {Yang}, \emph{\apj} \textbf{678},
  1316--1323 (2008). %, \eprint{arXiv:0801.0909}.

\bibitem[{Galloway} et~al.(2005)]{GM05}
D.~K. {Galloway}, C.~B. {Markwardt}, E.~H. {Morgan}, D.~{Chakrabarty}, and
  T.~E. {Strohmayer}, \emph{\apjl} \textbf{622}, L45--L48 (2005).

\bibitem[{Falanga} and {Titarchuk}(2007)]{FT07}
M.~{Falanga}, and L.~{Titarchuk}, \emph{\apj} \textbf{661}, 1084--1088 (2007). %,  \eprint{arXiv:astro-ph/0702453}.

\bibitem[{Pechenick} et~al.(1983)]{PFC83}
K.~R. {Pechenick}, C.~{Ftaclas}, and J.~M. {Cohen}, \emph{\apj} \textbf{274},
  846--857 (1983).

\bibitem[{Riffert} and {Meszaros}(1988)]{RM88}
H.~{Riffert}, and P.~{Meszaros}, \emph{\apj} \textbf{325}, 207--217 (1988).

\bibitem[{Page}(1995)]{Page95}
D.~{Page}, \emph{\apj} \textbf{442}, 273--285 (1995).

\bibitem[{Leahy} and {Li}(1995)]{LL95}
D.~A. {Leahy}, and L.~{Li}, \emph{\mnras} \textbf{277}, 1177--1184 (1995).

\bibitem[{Zavlin} et~al.(1995)]{ZSP95}
V.~E. {Zavlin}, Y.~A. {Shibanov}, and G.~G. {Pavlov}, \emph{Astronomy Letters}
  \textbf{21}, 149--158 (1995).

\bibitem[{Beloborodov}(2002)]{B02}
A.~M. {Beloborodov}, \emph{\apjl} \textbf{566}, L85--L88 (2002).

\bibitem[{Braje} et~al.(2000)]{BRR00}
T.~M. {Braje}, R.~W. {Romani}, and K.~P. {Rauch}, \emph{\apj} \textbf{531},
  447--452 (2000).

\bibitem[{Poutanen} and {Beloborodov}(2006)]{PB06}
J.~{Poutanen}, and A.~M. {Beloborodov}, \emph{\mnras} \textbf{373}, 836--844
  (2006). %, \eprint{arXiv:astro-ph/0608663}.

\bibitem[{Weinberg} et~al.(2001)]{WML01}
N.~{Weinberg}, M.~C. {Miller}, and D.~Q. {Lamb}, \emph{\apj} \textbf{546},
  1098--1106 (2001).

\bibitem[{Muno} et~al.(2002)]{MOC02}
M.~P. {Muno}, F.~{{\"O}zel}, and D.~{Chakrabarty}, \emph{\apj} \textbf{581},
  550--561 (2002).

\bibitem[{Poutanen}(2004)]{P04}
J.~{Poutanen}, 
%\enquote{{The Physics of X-ray Emission from Accreting Millisecond Pulsars},} 
in \emph{AIP Conf. Proc. 714: X-ray Timing 2003: Rossi and Beyond}, 
edited by P.~{Kaaret}, F.~K. {Lamb}, and J.~H. {Swank}, 2004,  pp. 228--231.

\bibitem[{Cadeau} et~al.(2005)]{CLM05}
C.~{Cadeau}, D.~A. {Leahy}, and S.~M. {Morsink}, \emph{\apj} \textbf{618},
  451--462 (2005).

\bibitem[{Kulkarni} and {Romanova}(2005)]{KR05}
A.~K. {Kulkarni}, and M.~M. {Romanova}, \emph{\apj} \textbf{633}, 349--357
  (2005).

\bibitem[{Cadeau} et~al.(2007)]{CMLC07}
C.~{Cadeau}, S.~M. {Morsink}, D.~{Leahy}, and S.~S. {Campbell}, \emph{\apj}
  \textbf{654}, 458--469 (2007). %, \eprint{arXiv:astro-ph/0609325}.

\bibitem[{Morsink} et~al.(2007)]{MLCB07}
S.~M. {Morsink}, D.~A. {Leahy}, C.~{Cadeau}, and J.~{Braga}, \emph{\apj}
  \textbf{663}, 1244--1251 (2007). %, \eprint{arXiv:astro-ph/0703123}.

\bibitem[{Viironen} and {Poutanen}(2004)]{VP04}
K.~{Viironen}, and J.~{Poutanen}, \emph{\aap} \textbf{426}, 985--997 (2004).

\bibitem[{Chandrasekhar}(1960)]{Cha60}
S.~{Chandrasekhar}, \emph{{Radiative transfer}}, New York: Dover, 1960.

\bibitem[{Sobolev}(1963)]{Sob63}
V.~V. {Sobolev}, \emph{{A treatise on radiative transfer}}, Princeton: Van
  Nostrand, 1963.

\bibitem[{Strohmayer} et~al.(2003)]{SM03}
T.~E. {Strohmayer}, C.~B. {Markwardt}, J.~H. {Swank}, and J.~{in't Zand},
  \emph{\apjl} \textbf{596}, L67--L70 (2003).

\bibitem[{Ozel}(2008)]{ozel08}
F.~{Ozel}, \emph{\apj}, in press (2008), \eprint{arXiv:0809.0509}.

\bibitem[{Lamb} et~al.(2008)]{lamb08}
F.~K. {Lamb}, S.~{Boutloukos}, S.~{Van Wassenhove}, et al., 
%R.~T. {Chamberlain}, K.~H. {Lo}, A.~{Clare}, W.~{Yu}, and M.~C. {Miller}, 
\emph{\apj}, submitted (2008), \eprint{arXiv:0808.4159}.

\bibitem[{Markwardt} et~al.(2002)]{MS02}
C.~B. {Markwardt}, J.~H. {Swank}, T.~E. {Strohmayer}, J.~J.~M.~i. {Zand}, and
  F.~E. {Marshall}, \emph{\apjl} \textbf{575}, L21--L24 (2002).

\bibitem[{Leahy} et~al.(2008{\natexlab{a}})]{LMC08}
D.~A. {Leahy}, S.~M. {Morsink}, and C.~{Cadeau}, \emph{\apj} \textbf{672},
  1119--1126 (2008{\natexlab{a}}). %, \eprint{arXiv:astro-ph/0703287}.

\bibitem[{Leahy} et~al.(2008{\natexlab{b}})]{LMCC08}
D.~A. {Leahy}, S.~M. {Morsink}, Y.-Y. {Chung}, and Y.~{Chou}, \emph{\apj},
  submitted (2008{\natexlab{b}}), \eprint{arXiv:0806.0824}.

\end{thebibliography}

\end{document}